\documentclass[journal]{IEEEtran}

\usepackage{cite}
\usepackage{amsmath,amssymb,amsfonts}
\usepackage{mathtools}
\usepackage{amsthm}
\usepackage{algorithmic}
\usepackage{graphicx}
\usepackage{textcomp}
\usepackage[protrusion=false,expansion=true]{microtype}
\usepackage{xcolor}
\usepackage{booktabs}
\usepackage{tabularx}
\usepackage{balance}
\usepackage{enumerate}
\usepackage{stfloats}
\usepackage{placeins}
\usepackage{mathrsfs}
\usepackage{etoolbox}
\usepackage[caption=false,font=footnotesize]{subfig}

\newtheorem{definition}{Definition}

\makeatletter
\renewcommand{\subsection}{%
  \@startsection{subsection}{2}{\z@}%
  {1.4ex}%
  {0.8ex}%
  {\normalfont\normalsize\itshape}%
}
\makeatother

\usepackage[
    bookmarks=false,
    colorlinks=true,
    citecolor=blue,
    linkcolor=blue,
    urlcolor=blue
]{hyperref}

\usepackage[capitalize]{cleveref}

\graphicspath{{img/}}

\AtBeginDocument{
    \setlength{\abovedisplayskip}{3pt plus 1pt minus 1pt}
    \setlength{\belowdisplayskip}{3pt plus 1pt minus 1pt}
    \setlength{\abovedisplayshortskip}{2pt plus 1pt minus 1pt}
    \setlength{\belowdisplayshortskip}{2pt plus 1pt minus 1pt}
}

\AtBeginEnvironment{thebibliography}{
    \sloppy
    \setlength{\itemsep}{-0.6pt}
    \setlength{\parsep}{0pt}
    \setlength{\parskip}{0pt}
    \renewcommand{\baselinestretch}{0.97}\selectfont
}

\begin{document}

\title{Reliability Analysis for Ground-Air Cooperative Wireless-Powered Internet-of-Things}

\author{Xingyu Huang,
        Zhican Zhong,
        Yulei Wang, 
        Yalin Liu,
        Yongqiang Cui, 
        and Di Bai

\thanks{The work is supported in part by the Hubei Provincial Natural Science Foundation of China under Grant No. 2026AFB282, in part by the Fundamental Research Funds for the Central Universities, South-Central Minzu University under Grant No. CZQ25005, in part by the Research Grants Council of the Hong Kong Special Administrative Region, China, under Project UGC/FDS16/E15/24, in part by the Hong Kong Metropolitan University Research Grant under Project PFDS/2025/34, and in part by the Young Scientific and Technological Talents Training Program of Hubei Province under Grant No. 2025DJA072. \textit{(Corresponding author: Yulei Wang, Yongqiang Cui.)}}

\thanks{X. Huang, Z. Zhong, Y. Wang, Y. Cui and D. Bai are with the Hubei Key Laboratory of Intelligent Wireless Communications, Hubei Engineering Research Center of Intelligent Internet of Things Technology, College of Electronics and Information Engineering (School of Robotics), South-Central Minzu University, Wuhan 430074, China (e-mail: \{xyhuang, zczhong, ylwang, cuiyq\}@mail.scuec.edu.cn, dibai@scuec.edu.cn).}

\thanks{Y. Liu is with the School of Science and Technology, Hong Kong Metropolitan University, Hong Kong (e-mail: ylliu@hkmu.edu.hk).}

\vspace{-8 mm}
}

\maketitle
\pagestyle{empty}       
\thispagestyle{empty}

\begin{abstract}
Battery-constrained Internet of Things (IoT) devices face significant challenges in maintaining long-term uplink data collection, especially in network-limited scenarios. This paper investigates a Ground-Air cooperative Wireless-Powered IoT network (GA-WPIoT), where ground power beacons (GPBs) provide radio-frequency energy supply and unmanned aerial vehicles (UAVs) serve as flexible aerial access points for IoT data collection. Unlike conventional UAV-enabled wireless-powered solutions, GA-WPIoT decouples energy transfer and data collection, reducing the hardware and energy burden of UAVs. 
To capture the device-level performance variations among IoT devices, we develop a stochastic geometry-based analytical framework, where UAVs are modeled by a Mat\'ern hard-core point process, IoT devices by a Poisson point process, and GPBs by a finite annular binomial point process. The conditional success probabilities of the GPB-device link (for energy transfer) and the device-UAV link (for data collection) are characterized, and device-level reliability distributions are obtained through the Meta-Distribution based analysis. Numerical results under suburban and urban environments reveal the impacts of UAV altitude and annular GPB deployment on reliability distributions, providing useful guidelines for GA-WPIoT.
\vspace{-2mm}
\end{abstract}

\begin{IEEEkeywords}
UAV communications, wireless energy transfer, meta distribution, ground power beacon, stochastic geometry
\vspace{-3mm}
\end{IEEEkeywords}

\section{Introduction}
Internet of Things (IoT) applications are proliferating by continuously collecting massive and heterogeneous data to support various data-driven services, e.g., smart cities, ocean monitoring, and intelligent agriculture\cite{Ru2026SG}. For low-cost deployments, these applications are typically implemented using battery-constrained IoT devices under network-limited deployment scenarios, which leads to a critical sustainability challenge in maintaining long-term and stable data collection. Instead of establishing costly network infrastructures, the integration of ground power beacons (GPBs) and unmanned aerial vehicles (UAVs) offers a low-cost and on-demand solution for wireless-powered IoT data collection. For brevity, we refer to this solution as a Ground-Air cooperative Wireless-Powered IoT network (GA-WPIoT). In GA-WPIoT, multiple GPBs are deployed to power distributed IoT devices via radio-frequency (RF) energy transfer\cite{Talgat2024uplink}, while UAVs act as aerial access points that provide flexible and on-demand coverage to collect uplink data from ground IoT devices~\cite{Liu23UAV,wei2022UAV}. Unlike conventional UAV-enabled wireless-powered solutions, GA-WPIoT decouples energy transfer and data collection, reducing the hardware and energy burden of UAVs. 

Analyzing the reliability of GA-WPIoT overall IoT devices is crucial for its practical design and realization, yet it is challenging for two main reasons. On the one hand, conventional spatially averaged network metrics are not suitable to capture the reliability of data collection performance over different IoT devices in GA-WPIoT~\cite{monemi2025higher}. Given a specific IoT device, the spatial deployment of its nearby GPBs together with the serving UAV jointly determines the data collection performance, resulting in substantial device-level performance variations across different network realizations. On the other hand, successful data collection must satisfy a coupled energy-communication condition: each IoT device must first harvest sufficient energy from nearby GPBs and then reliably complete uplink transmission to its serving UAV. This coupled success condition complicates system performance analysis. Devices located far from GPBs may suffer from insufficient harvested energy, whereas devices situated near the edge of UAV service regions may experience severe device-UAV path loss and stronger uplink interference~\cite{Zhou2022PB}. To sum up, in consideration of spatial deployments of GPBs, UAVs, and the IoT devices themselves, a device-level reliability analysis method is required to characterize the performance of GA-WPIoT.

Prior studies on existing wireless-powered communication networks have analyzed the energy harvesting
\cite{Zhao26Reconfigurable,Zhao26FASWIPT}
and uplink transmission performance, but typically under oversimplified spatial models, such as homogeneous user distributions or fixed GPB deployments~\cite{Talgat2024uplink}.
UAV-assisted IoT networks have been investigated mainly by focusing on UAV trajectory design and communication performance, often assuming battery-powered devices or neglecting the spatial characteristics of energy provisioning~\cite{Liu23UAV, wei2022UAV}. Overall, the joint spatial effects of GPB deployment, UAV spatial distribution, and random IoT-device locations in a ground-air cooperative wireless-powered IoT network, i.e., GA-WPIoT, have not been thoroughly characterized~\cite{liu2025unified,Wang25SAGIN}. To capture the device-level reliability, the meta distribution (MD) is a suitable tool to analyze data collection performance by characterizing the conditional success probability (CSP) of each IoT device, as well as its distribution across the spatial distribution of the served IoT device~\cite{Wang24WiFi, qin2023downlink}. Whereas~\cite{Wang24WiFi, qin2023downlink} studies clustered power-beacon deployment and meta distributions for energy and information transfer using a truncated Poisson cluster model, our model combines a fixed number of GPBs in each finite annulus with spatially
separated aerial collectors modeled by a Type-II MHCPP. Relative to the UAV-assisted downlink SINR analysis in~\cite{Wang24WiFi}, we characterize the joint energy-qualification/uplink-decoding reliability of a ground--air architecture in which GPBs supply energy and UAVs collect data.
To overcome this research gap, this paper has the following contributions. 
\begin{itemize}
    \item \textit{First,} we develop a stochastic-geometry model for GA-WPIoT by jointly modeling UAVs' deployment as a Mat\'ern hard-core point process, random IoT-device locations by a Poisson point process, and finite annular GPBs' placement by a binomial point process.
    \item \textit{Second,} we derive the moments of the CSP for the GPB-device energy-transfer link and the device-UAV link, accounting for finite GPB deployment and spatially distributed uplink interference. These results enable the device-level reliability of both links to be characterized through meta distributions.
    \item \textit{Third,} we obtain the meta distribution (MD) of the overall performance by jointly considering the harvested-power qualification and uplink-decoding requirements. Numerical results reveal the joint effects of the GPB inner radius, UAV altitude, and communication parameters, providing design guidelines for GA-WPIoT.
\end{itemize}

\vspace{-0.2cm}
\section{System Model}\label{sec:system-model}
\subsection{Network Model}
\begin{figure}[!t]
\centering
\includegraphics[width=0.9\columnwidth,height=0.5\textheight,keepaspectratio]{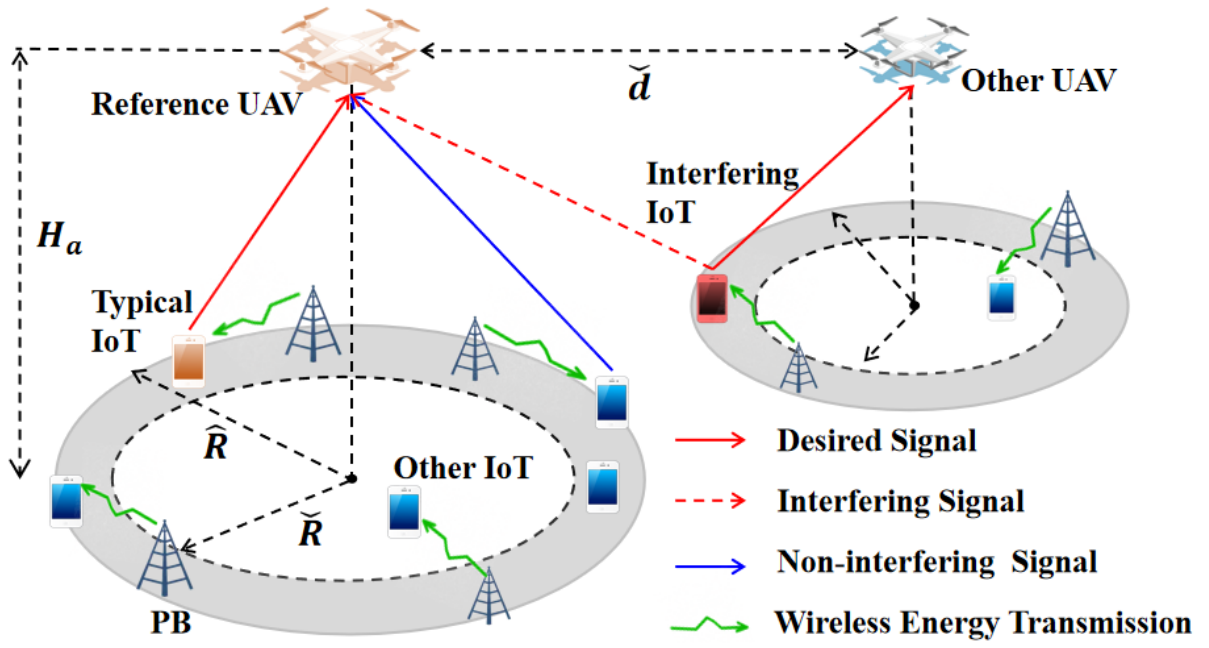}
\caption{An overview of GA-WPIoT.}
\label{Fig:system_model}
\end{figure}

\cref{Fig:system_model} shows a typical GA-WPIoT, where multiple UAVs hover at a fixed altitude $H_a$ and act as aerial access points to collect uplink information from ground IoT devices. Meanwhile, ground power beacons (GPBs) are deployed to provide RF energy for energy-constrained IoT devices. Therefore, the network consists of two constituent links: i) the GPB-device link for energy transfer, and ii) the device-UAV link for uplink transmission. To capture the spatial separation among UAVs, the locations of UAVs are modeled by a Type-II Matérn hard-core point process (MHCPP) $\Phi_a$ with density $\lambda_a$~\cite{Wang25SAGIN, Dong2024SG}. Specifically, $\Phi_a$ is obtained by thinning a parent Poisson point process (PPP) $\Phi_a^0$ with density $\lambda_a^0$ and hard-core distance $\check d$. The density of $\Phi_a$ is given by $\lambda_a=\lambda^{0}_{a} p_a(\check{d})$, where $p_a(\check{d})$ is the retention probability of each point in $\Phi_a^0$ and given by
\begin{equation}
p_a(\check{d}) = \begin{cases}
\dfrac{1-\exp\left(-\lambda^{0}_{a} \pi\check{d}^2\right)}
{\lambda^{0}_{a}\pi \check{d}^2}, & \check{d}>0, \\[1.5ex]
1, & \check{d}=0.
\end{cases}
\end{equation}
For analytical tractability, the service region of each UAV is approximated by a disk with radius $\hat{R}$, i.e., $\mathcal A_u = \left\{ \mathbf x\in\mathbb R^2: \|\mathbf x\| \le \hat{R} \right\}$, where $\|\mathbf x\|$ is the Euclidean distance between $\mathbf x$ and the ground projection of its associated UAV. Only IoT devices located within the service region of their associated UAVs are allowed to establish uplink links.

The IoT devices within $\mathcal A_u$ are modeled by an independent PPP $\Phi_u$ with density $\lambda_u$. To improve the energy transmission for edge IoT devices, GPBs are deployed in an annular region centered at each UAV. Since the number of GPBs is usually predetermined by deployment planning, the GPB locations are modeled by a binomial point process (BPP) $\Phi_p$ \cite{Talgat2024uplink} consisting of $N_p$ independently and uniformly distributed GPBs within $\mathcal A_{\rm P} = \left\{\mathbf y\in\mathbb R^2: \check{R} \le \|\mathbf y\| \le \hat{R} \right\}$. The annular deployment provides controllable spatial energy provisioning. 

\vspace{-2mm}
\subsection{Link Model}
Both the GPB-device link (for energy transfer) and the device-UAV link (for uplink transmission) experience small-scale fading and large-scale path loss. Let $i\in\{1,2\}$ denote the GPB-device and device-UAV links, respectively. The small-scale fading power gain of the $i$-th link is denoted by $h_i$. Rayleigh fading is assumed for both links, thus, $h_i$ follows an exponential distribution with unit mean, i.e, $h_i\sim\exp(1)$\cite{Talgat2024uplink}. 

\subsubsection{GPB-device Link} 
We assume that each IoT device harvests RF energy from the nearest GPB deployed in the annular region of its associated UAV. Let $d_1$ denote the distance between the typical IoT device and its nearest GPB. Since both GPBs and IoT devices are located on the ground, the large-scale path loss is modeled by $d_1^{-\alpha_1}$, where $\alpha_1$ denotes the path-loss exponent of the GPB-device link. Thus, the instantaneous harvested power $P_{\rm harv}$ at the typical IoT device is given by
\begin{equation}
P_{\rm harv} = \eta G P_{\rm G} h_1 d_1^{-\alpha_1},
\label{energy_1}
\end{equation}
where $\eta \in (0,1]$ denotes the radio-frequency-to-direct-current energy conversion efficiency, $G$ is the directional antenna gain of the GPB, and $P_{\rm G}$ is the transmit power of each GPB.

\subsubsection{Device-UAV Link} 
Without loss of generality, the device-UAV link experiences a probabilistic line-of-sight/non-line-of-sight (LoS/NLoS) path-loss model to capture the air-to-ground propagation characteristics. Let $r_u$ denote the horizontal distance between the IoT device and the ground projection of its associated UAV. The corresponding device-UAV communication link distance is given by $d_2(r_u) = \sqrt{r_u^2+H_a^2}$. Let $g_2(r_u)$ denote the large-scale channel power gain of the device-UAV link, which is given by~\cite{Liu23UAV}
\begin{equation}
g_2(r_u)=10^{-\frac{P_{\rm L}(r_u)L_{\rm L}(r_u)+P_{\rm N}(r_u)L_{\rm N}(r_u)}{10}},
\label{eq:big path loss}
\end{equation}
where $L_{j}(r_u) (j \in \{\rm L, N\})$ denotes the LoS and NLoS path losses (in dB) with their corresponding probabilities $\mathcal{P} _{j}(r_u)$ (i.e. $\mathcal{P}_{\rm N}(r_u) + \mathcal {P}_{\rm L}(r_u) = 1$), respectively. $L_{j}(r_u)$ is given by
\begin{equation}
L_j(r_u)=L_{\rm FS}(d_0)+10\alpha_j\log_{10}\left(\frac{d_2(r_u)}{d_0}\right),
\end{equation}
where $\alpha_j$ denotes the path-loss exponents of the LoS and NLoS components, respectively. $L_{\rm FS}(d_0)$ is the free-space path loss with $d_0$ being the reference distance, and is given by
\begin{equation}
L_{\rm FS}(d_0)=20\log_{10}\left(\frac{4\pi f_2 d_0}{c}\right),
\end{equation}
where $f_2$ is the carrier frequency of the device-UAV link and $c$ is the speed of light. Besides, $\mathcal{P}_{\rm L}(r_u)$ is given by\cite{al2014optimal}
\begin{equation}
\mathcal{P}_{\rm L}(r_u)=\frac{1}{1+\chi \exp\left[-\xi\left(\theta(r_u)-\chi\right)\right]},
\end{equation}
where $\chi$ and $\xi$ are environment-dependent parameters\cite{al2014optimal}, $\theta(r_u)$ is the elevation angle from the IoT device to the UAV and given by 
\begin{equation}
\theta(r_u) = \frac{180}{\pi} \arcsin \left( \frac{H_a}{d_2(r_u)} \right) \ (\rm in \ degrees).
\end{equation}

\vspace{0mm}
\subsection{Interference and SINR Model}\label{sec:Interference_SINR}
The uplink transmission adopts the orthogonal frequency division multiple access scheme to serve multiple IoT devices. Specifically, each resource block is allocated to at most one IoT device in each UAV service region. Therefore, intra-cell interference is eliminated, while co-channel interference is generated by IoT devices located in neighboring UAV service regions and reusing the same resource block. 

An IoT device is regarded as active only if $P_{\rm harv}\ge P_{\min}$, where $P_{\min}$ denotes a minimum harvested-power threshold for activating uplink transmission with channel inversion power control. Under a fully-loaded scheduling assumption, each neighboring UAV service region schedules one active co-channel IoT device on the considered resource block. Hence, the density of active interfering IoT devices is approximated by the density of UAVs~\footnote{Since the probability generating functional (PGFL) of the MHCPP is generally intractable, the active co-channel interfering process is approximated by an equivalent PPP with density $\lambda_u^{\rm I}$ while preserving an exclusion radius of $2\hat R$ around the typical UAV.}, i.e.,
\begin{equation}
\lambda_u^{\rm I} \approx \lambda_a .
\end{equation} 

For active IoT devices, large-scale channel inversion power control is adopted~\cite{Zhou2022PB}. This compensates for the distance-dependent attenuation of the device-UAV link such that the average received power at the serving UAV is maintained at a target value $\rho_u$. The SINR at the typical UAV is given by
\begin{equation}
{\rm SINR}
=\frac{P_r}{I+W},
\label{SINR}
\end{equation}
where $W$ is an additive white Gaussian noise.
According to channel inversion power control, the desired signal power is
\begin{equation}
P_r=\rho_u h_2^0.
\end{equation}
For the $k$-th active interfering IoT device $\mathbf x_k\in\Phi_u^{\rm I}$, let $r_u^k$ denote the horizontal distance between the interfering IoT device and its serving UAV projection, and let $r_u^{k'}$ denote the horizontal distance between the interfering IoT device and the typical UAV projection. The transmit power of the interfering IoT device is
\begin{equation}
P_k^{\rm tx}
=\frac{\rho_u}{g_2(r_u^k)}.
\end{equation}

Therefore, the aggregate interference is
\begin{equation}
I=\sum_{\mathbf x_k\in\Phi_u^{\rm I}}\rho_u h_2^{k'}\frac{g_2(r_u^{k'})}{g_2(r_u^k)},
\end{equation}
where $h_2^{k'}$ denotes the small-scale fading power gain.
\begin{figure}[!t]
\centering
\includegraphics[width=0.75\columnwidth]{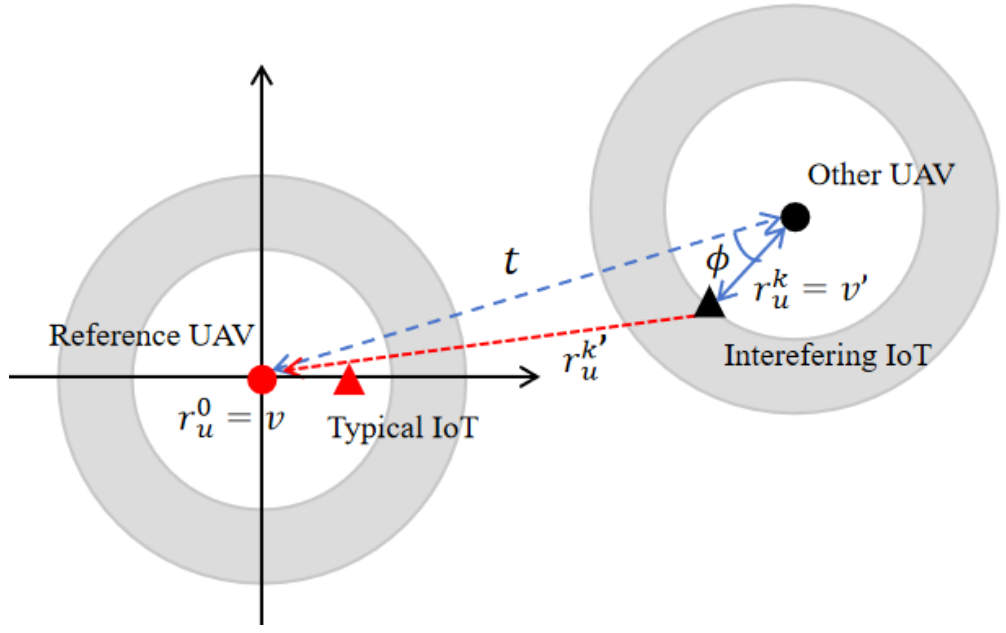}
\caption{Illustration of the uplink interference geometry.}
\label{Fig-Intefer}
\end{figure}

\vspace{0mm}
\section{Performance Analysis}
In this section, we first define the success of a data collection in GA-WPIoT, the CSP, and the CSP's MD to characterize the device-level reliability across the network variations. Accordingly, we further analyze and derive the device-level reliability (i.e., the CSP's MD) for the GPB-device link, the device-UAV link, and both links, respectively.

\begin{definition}
A data collection is successful if two conditions are satisfied: i) the harvested power at the IoT device exceeds a threshold $P_{\min}$, and ii) the received SINR at its associated UAV exceeds a threshold $\gamma_{\mathrm{th}}$ (to complete uplink decoding). 
\end{definition}

\begin{definition}
Given a specific IoT device, its CSP on a GPB-device link or a device-UAV link or both links is defined as the probability that the harvested power or the received SINR or both requirements are satisfied, averaged over small-scale fading and distributed network nodes. 
\end{definition}

\begin{definition}
Given a GPB-device link or a device-UAV link or both links, the device-level reliability is characterized by the MD $\bar F_{\mathcal{P}_s}(x)$ of its CSP $\mathcal{P}_s$. By calculating the fraction of CSP values that exceed a prescribed reliability target $x$, the MD-characterized reliability $\bar F_{\mathcal{P}_s}(x)$ is determined by its complementary cumulative distribution function, i.e.,
\begin{equation}
\bar F_{\mathcal{P}_s}(x) \triangleq \mathbb{P}(\mathcal{P}_s>x), \quad 0\leq x\leq 1.
\end{equation}
For calculation tractability, we approximate the above CSP's MD with a standard beta distribution with respect to the CSP's first two moments~\cite{Wang24WiFi}, which yields
\begin{equation}
\bar F_{\mathcal{P}_s}(x)\approx 1-I_x(a,b),
\label{meta1}
\end{equation}
where $I_x(a,b)$ is the regularized incomplete beta function, and
\begin{equation}
a= \frac{M_1(M_1-M_2)}{M_2-M_1^2}, \quad b= \frac{(1-M_1)(M_1-M_2)} {M_2-M_1^2}.
\label{meta2}
\end{equation}
where $M_b=\mathbb{E}[\mathcal{P}_s^b]$ denote the $b$-th moment of $\mathcal{P}_s$.
\end{definition}

\subsection{Reliability on the GPB-device Link}
\label{sec-PB_IoT_energy}
Here, we analyze the CSP's MD on the GPB-device link. Let $M_b^{\rm I}$ denote the $b$-th moment of the GPB-device-link CSP, 
which is calculated by 
\begin{equation}\label{Eq-Mb1-Mean}
M_b^{\rm I} = \mathbb{E}_{r_u^0}\left[M_b^{\rm I}(v)\right] 
= \int_{0}^{\hat R} M_b^{\rm I}(v) f_{r_u^0}(v)\mathrm{d}v ,
\end{equation}
where $M_b^{\rm I}(v)$ denotes the conditional $b$-th moment of the CSP given $r_u^0=v$, and $f_{r_u^0}(v)$ is the probability density function (PDF) of $r_u^0$. Since the typical IoT device is uniformly distributed in the UAV service disk, $f_{r_u^0}(v)$ is given by $f_{r_u^0}(v)=2v/\hat R^2$, $0\leq v\leq \hat R$. 

Besides, $M_b^{\rm I}(v)$ is given by
\begin{equation}\label{Eq-Mb1-Conditional}
M_b^{\rm I}(v) = \mathbb{E}_{d_1^0}\left[ \left( \mathcal{P}_1(d_1^0) \right)^b \right] 
= \int_{0}^{\hat R+v} \left( \mathcal{P}_1(r)\right)^b f_{d_1^0}(r|v)\mathrm{d}r,
\end{equation}
where $d_1^0 =\min_{1\leq i\leq N_p}d_1^i$ is the distance from the typical IoT device to the nearest GPB, which is the minimum distance from the typical IoT device to the $i$-th GPB, $\mathcal{P}_1(r)$ denotes the CSP conditioned on the realization $d_1^0 = r$. and $f_{d_1^0}(r|v)$ is the conditional PDF of $d_1^0$ given $r_u^0=v$. Specifically, $f_{d_1^0}(r|v)$ is obtained by differentiating its conditional CDF $F_{d_1^0}(r|v)$, i.e., $f_{d_1^0}(r|v) = \partial F_{d_1^0}(r|v)/\partial r$, where $F_{d_1^0}(r|v)$ is determined by the disk-annulus intersection geometry. The corresponding CDF and PDF are provided in Appendix~\ref{app:nearest_gpb}.

\noindent Given $d_1^0 = r$, the CSP of the GPB-device link $\mathcal{P}_1(r)$ is
\begin{equation}\label{Eq-CSP-P1}
\begin{aligned}
    &\mathcal{P}_1(r) = \mathbb{P}(P_{\rm harv}\geq P_{\min} \rvert d_1^0 = r) \\
    & = \mathbb{P}\left(h_1 \geq \frac{P_{\min} r^{\alpha_1}} {\eta G P_{\rm G}} \rvert d_1^0 = r \right) = \exp\left(-\beta r^{\alpha_1}\right),
\end{aligned}
\end{equation}
where $\beta=P_{\min}/(\eta G P_{\rm G})$, and the last equality follows from $h_1\sim\exp(1)$.

Consequently, substituting~\eqref{Eq-CSP-P1} into \eqref{Eq-Mb1-Conditional}, we obtain the conditional b-th moment of the CSP and subsitiute \eqref{Eq-Mb1-Conditional} into \eqref{Eq-Mb1-Mean}, we obtain  the b-th moment~\eqref{Eq-Mb1-Mean} gives
\begin{equation}
M_b^{\rm I} = \int_{0}^{\hat R} \int_{0}^{\hat R+v} \exp\left(-b\beta r^{\alpha_1}\right) f_{d_1^0}(r|v) \frac{2v}{\hat R^2}
\mathrm{d}r\mathrm{d}v .
\end{equation}
The first two moments $M_1^{\rm I}$ and $M_2^{\rm I}$ are obtained by setting $b=1$ and $b=2$, respectively. Then, MD of the GPB-device link, denoted by $\bar F_1(x)$, is obtained from the beta approximation in \eqref{meta1} and \eqref{meta2} by setting $M_1=M_1^{\rm I}$ and $M_2=M_2^{\rm I}$.

\subsection{Reliability of the device-UAV link}

Here, we analyze MD of the device-UAV uplink. Let $M_b^{\rm U}$ denote the $b$-th moment of the uplink CSP $\mathcal{P}_2$. Since $\mathcal{P}_2$ depends on the active interfering IoT process $\Phi_u^{\rm I}$ and the random active-IoT locations, $M_b^{\rm U}$ is given by
\begin{equation}
\begin{aligned}
M_b^{\rm U}
&=\mathbb{E}_{\Phi_u^{\rm I}, h_2^{k'}}\left[ \left(\mathcal{P}_2\right)^b\right].
\end{aligned}
\label{Eq-Mb2-Mean}
\end{equation}
In~\eqref{Eq-Mb2-Mean}, the CSP of $\mathcal{P}_2$ is given by
\begin{equation}\label{Eq-CSP-P2}
\begin{aligned}
\mathcal{P}_2 
&= \mathbb{P}({\rm SINR}>\gamma_{\rm th}) = \mathbb{P}\left(h_2^0>\frac{\gamma_{\rm th}(I+W)}{\rho_u}|\Phi_u^{\rm I}, h_2^{k'} \right) \\
&= \exp\left(-\frac{\gamma_{\rm th}W}{\rho_u}\right) \mathbb{E}_I \left[\exp\left(-\frac{\gamma_{\rm th}I }{\rho_u}\right)\right], \\
&= \exp\left(-\frac{\gamma_{\rm th}W}{\rho_u}\right)\mathcal{L}_{\Phi_u^{\rm I},h_2^{k'}}\left(\frac{\gamma_{\rm th}}{\rho_u} \right),
\end{aligned}
\end{equation}
where $\mathcal{L}_{\Phi_u^{\rm I},h_2^{k'}}(s)=\mathbb{E}_{\Phi_u^{\rm I},h_2^{k'}}[\exp(-sI)]$ denotes the Laplace transform of the aggregate interference, and the last equality follows from $h_2^0\sim\exp(1)$.

For compactness, let $\chi(t,v',\phi)\triangleq g_2(r_u^{k'})/g_2(v')$ with $r_u^{k'}=\sqrt{t^2+(v')^2-2tv'\cos\phi}$. Using the aggregate interference expression in Section~\eqref{sec:Interference_SINR}, the Laplace transform term is derived as
\begin{equation}
\label{eq:laplace_interference}
{\footnotesize
\begin{aligned}
& \mathcal{L}_{\Phi_u^{\rm I},h_2^{k'}}\left(\frac{\gamma_{\rm th}}{\rho_u}\right)\overset{\phantom{(a)}}{=}\mathbb{E}_{\Phi_u^{\rm I}}
\left[\mathbb{E}_{h_2^{k'}}\left[\exp\left(-\gamma_{\rm th}\sum_{\mathbf{x}_k\in\Phi_u^{\rm I}}h_2^{k'}\frac{g_2(r_u^{k'})}{g_2(r_u^k)}\right)\right]\right]\\& \overset{(a)}{=}\mathbb{E}_{\Phi_u^{\rm I}}\left[\prod_{\mathbf{x}_k\in\Phi_u^{\rm I}}\mathbb{E}_{r_u^k,\phi}\left[\frac{1}{1+\gamma_{\rm th}\chi(t,v',\phi)}
\right]\right]\\& \overset{(b)}{=}\exp\Bigg(2\pi\lambda_u^{\rm I}\int_{2\hat R}^{\infty}\Bigg[\mathbb{E}_{r_u^k,\phi}\left[\frac{1}{1+\gamma_{\rm th}\chi(t,v',\phi)}\right]-1\Bigg]t\,\mathrm{d}t\Bigg)\\& \overset{(c)}{=}\exp\Bigg(2\pi\lambda_u^{\rm I}\int_{2\hat R}^{\infty}\Bigg[\int_{0}^{\hat R}\int_{0}^{2\pi}\frac{1}{1+\gamma_{\rm th}\chi(t,v',\phi)}\\& \hspace{1.2cm}\times\frac{v'}{\hat \pi R^2}\mathrm{d}\phi\,\mathrm{d}v'
-1\Bigg]t\,\mathrm{d}t\Bigg),
\end{aligned}}
\end{equation}
where step $(a)$ follows from $h_2^{k'}\sim\exp(1)$ and the independence of the interfering fading gains. Step $(b)$ follows from the PGFL of an independently marked PPP~\cite{Afpc2018} with mark $\mathbf{m}=(r_u^k,\varphi)$, i.e.,
$\mathbb{E}[\prod_{\mathbf{x}\in\Phi}f(\mathbf{x},\mathbf{m})]
=\exp(-\lambda\int_{\mathbb{R}^2}[1-\mathbb{E}_{\mathbf{m}}\{f(\mathbf{x},\mathbf{m})\}] \mathrm{d}\mathbf{x})$, together with the polar-coordinate transformation and the exclusion radius $2\hat R$. Step $(c)$ expands the mark expectation over the serving-link distance $r_u^k$ and the angular location $\varphi$, where $t$ is the distance from an interfering UAV projection to the typical UAV projection, $v'$ is the realization of $r_u^k$, and $f_{r_u^k}(v')=2v'/\hat R^2$, $f_{\varphi}(\phi)=1/(2\pi)$. 
Accordingly, the $b$-th moment of the uplink CSP is obtained as
\begin{equation}
\label{eq:moment_u}
{\footnotesize
\begin{aligned}
M_b^{\rm U}&=\exp\Bigg(-\frac{b\gamma_{\rm th}W}{\rho_u}+2\pi\lambda_u^{\rm I}\int_{2\hat R}^{\infty}\Bigg[\int_{0}^{\hat R}\int_{0}^{2\pi}\frac{1}{\left(1+\gamma_{\rm th}\chi(t,v',\phi)\right)^b}\\&\hspace{3.4cm}\times\frac{v'}{\hat \pi R^2}
\mathrm{d}\phi\,\mathrm{d}v'-1\Bigg]t\,\mathrm{d}t\Bigg).
\end{aligned}}
\end{equation}

The first two moments $M_1^{\rm U}$ and $M_2^{\rm U}$ are obtained by setting $b=1$ and $b=2$, respectively. Then, MD of the device-UAV uplink, denoted by $\bar F_2(x)$, is obtained from the beta approximation by setting $M_1=M_1^{\rm U}$ and $M_2=M_2^{\rm U}$.

\subsection{Reliability of both links}

A successful device-UAV transmission in the considered network requires both successful GPB-device energy transfer and successful device-UAV uplink transmission. Since the energy-transfer and uplink fading processes are independent conditioned on the network geometry, the overall CSP is
\begin{equation}
P_{\rm cov}=P_1P_2,
\end{equation}
where $P_1$ and $P_2$ denote the CSPs of the GPB-device link and the device-UAV uplink, respectively.

For analytical tractability, $P_1$ and $P_2$ are further approximated as independent across network realizations. Then, the $b$-th moment of $P_{\rm cov}$ is\cite{qin2023downlink}
\begin{equation}
M_b^{\rm cov}
=\mathbb{E}\left[P_{\rm cov}^b\right]\approx M_b^{\rm I}M_b^{\rm U}.
\end{equation}
Accordingly, $M_1^{\rm cov}\approx M_1^{\rm I}M_1^{\rm U}$ and $M_2^{\rm cov}\approx M_2^{\rm I}M_2^{\rm U}$.

Using the beta approximation with $M_1^{\rm cov}$ and $M_2^{\rm cov}$, MD of the overall CSP, denoted by $\bar F_{\rm cov}(x)$, is obtained from \eqref{meta1} and \eqref{meta2} by setting $M_1=M_1^{\rm cov}$ and $M_2=M_2^{\rm cov}$.

\section{Numerical Results}

Unless otherwise stated, we set $\hat{R}=90\,\mathrm{m}$, $N_p=12$,$\lambda_a^0=5\times10^{-6}\, \mathrm{m}^{-2}$, $P_G=30\,\mathrm{dBm}$, $\alpha_1=3$, $\eta=0.8$, and $\rho_u=-67\,\mathrm{dBm}$. The LoS and NLoS path-loss exponents are $(\alpha_L,\alpha_N)=(2.1,2.4)$ ~\cite{Liu23UAV}. The environment parameters are $(\chi,\xi)=(4.88,0.43)$ for suburban environments and $(9.61,0.16)$ for urban environments\cite{al2014optimal}. Monte Carlo simulations use 3000 independent network realizations. In all figures, "Ana." and "Sim." denote analytical and simulation results, respectively.

\begin{figure}[!t]
    \raggedright
    \includegraphics[width=0.4\textwidth]{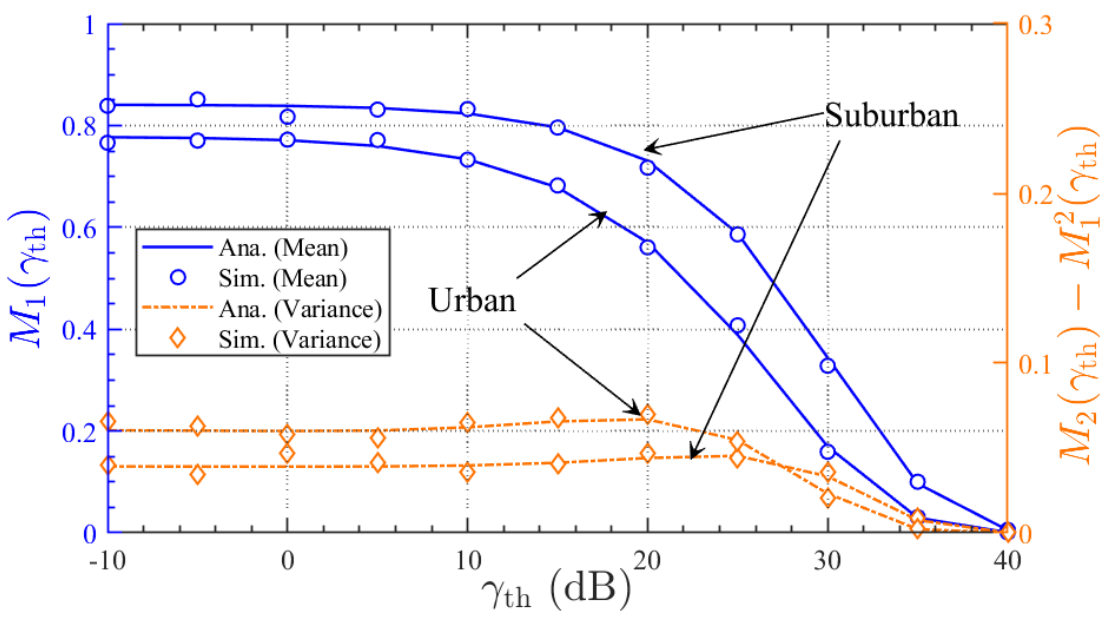}
    \caption{$M_1^{\rm cov}(\gamma_{\rm th})$ and $M_2^{\rm cov}(\gamma_{\rm th})-(M_1^{\rm cov}(\gamma_{\rm th}))^2$ vs. $\gamma_{\rm th}$.}
    \label{fig:average_variance}
\end{figure}

\begin{figure}[!t]
    \raggedright
    \includegraphics[width=0.38\textwidth]{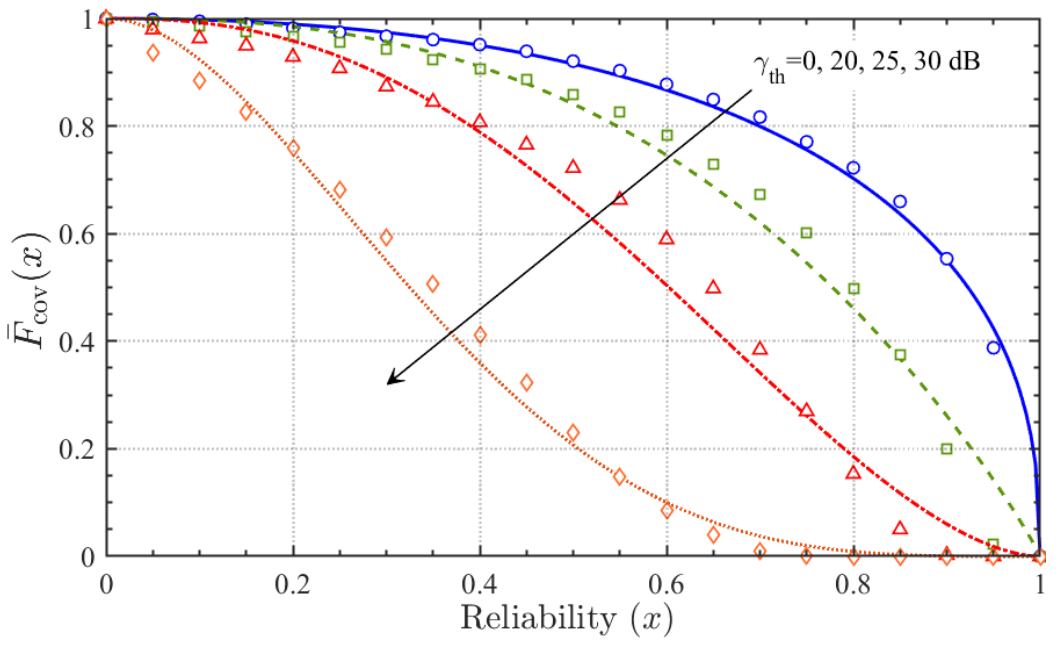}
    \caption{$\bar{F}_{\rm cov}(x)$ vs. reliability $x$ for different $\gamma_{\rm th}$ in the suburban environment.}
    \label{fig:meta_distribution_gamma}
\end{figure}

Fig.~\ref{fig:average_variance} shows $M_1^{\rm cov}(\gamma_{\rm th})$ and $M_2^{\rm cov}(\gamma_{\rm th})-(M_1^{\rm cov}(\gamma_{\rm th}))^2$ versus $\gamma_{\rm th}$ in suburban and urban environments. The analytical results match well with the simulation results, verifying the derived moments. As $\gamma_{\rm th}$ increases, $M_1^{\rm cov}(\gamma_{\rm th})$ decreases because a higher SINR threshold imposes a stricter decoding requirement. Compared with the urban environment, the suburban environment achieves a higher mean overall CSP due to more favorable air-to-ground propagation. Moreover, the variance first increases and then decreases with $\gamma_{\rm th}$, indicating that reliability heterogeneity among IoT devices is more evident in the medium-threshold region.

Fig.~\ref{fig:meta_distribution_gamma} $\bar{F}_{\rm cov}(x)$ versus the CSP target $x$ for different values of $\gamma_{\rm th}$ in the suburban environment. For a given $\gamma_{\rm th}$, $\bar{F}_{\rm cov}(x)$ decreases as $x$ increases because fewer IoT-device links achieve a higher CSP target. For a given $x$, a larger $\gamma_{\rm th}$ results in a lower $\bar{F}_{\rm cov}(x)$, indicating that a stricter uplink-decoding requirement reduces the fraction of IoT-device links whose overall CSP exceeds $x$. The close agreement between the analytical and simulation results confirms the accuracy of the beta approximation.

\begin{figure}[!t]
    \centering
    \includegraphics[width=1\columnwidth]{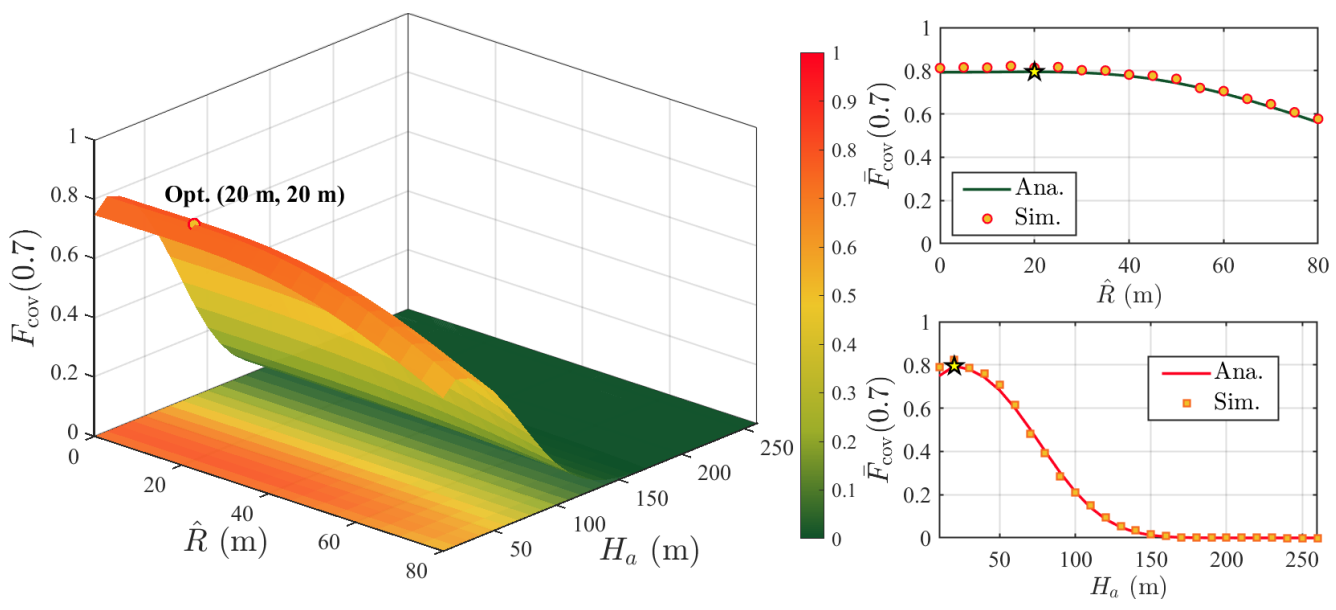}
    \vspace{-1mm}
    \centerline{\footnotesize (a) Suburban}
    \vspace{1mm}

    \includegraphics[width=0.98\columnwidth]{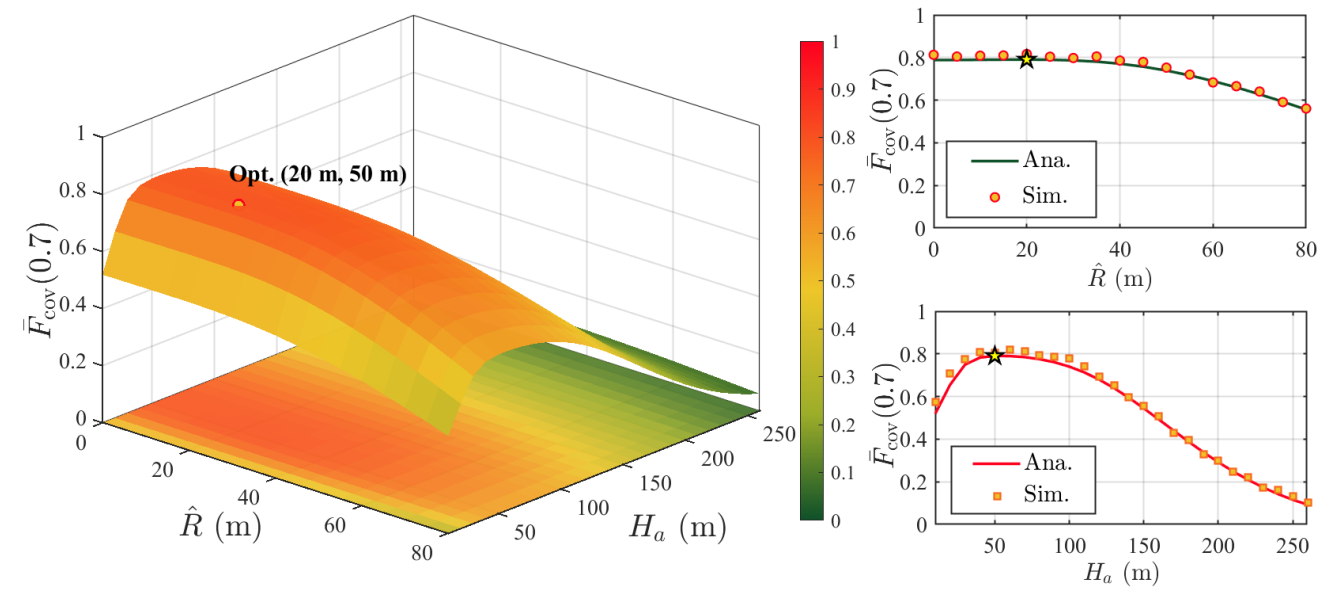}
    \vspace{-1mm}
    \centerline{\footnotesize (b) Urban}

    \vspace{0mm}
    \caption{$\bar{F}_{\mathrm{cov}}(0.7)$ vs. $\check{R}$ and $H_a$ in suburban and urban environments.}
    \label{Fig:heatmap_all}
    \vspace{0mm}
\end{figure}

Fig.~\ref{Fig:heatmap_all} illustrates $\bar{F}_{\rm cov}(0.7)$ versus the GPB inner radius $\check{R}$ and the UAV altitude $H_a$ in suburban and urban environments. In the suburban case, the optimal deployment is around $(\check{R},H_a)=(20~{\rm m},20~{\rm m})$, whereas in the urban case it shifts to approximately $(20~{\rm m},50~{\rm m})$. Compared with the suburban environment, the optimal UAV altitude becomes higher in urban areas because the improvement in the LoS probability outweighs the additional propagation loss caused by a longer transmission distance. Meanwhile, increasing the GPB inner radius improves the harvested energy of edge IoT devices by moving GPBs closer to the cell boundary. However, an excessively large inner radius weakens the energy supply for center IoT devices due to the increased GPB-device distance. These observations indicate that optimizing only the UAV altitude or only the GPB deployment is insufficient for maximizing the overall reliability. Instead, the UAV altitude and GPB deployment should be jointly designed according to the propagation environment to balance wireless energy provisioning and uplink transmission reliability.

\begin{figure*}[!t]
    \centering
    \includegraphics[width=0.9\linewidth]{Fig-Five-Cases.pdf}
    \caption{Illustration of the five geometric configurations for evaluating the intersection area
    $\left|\mathcal{B}(\mathbf{z}_0,r)\cap \mathcal{A}_{\rm P}\right|$.}
    \label{Fig:five_geometric_cases}
\end{figure*}

\vspace{-3 mm}
\section{Conclusion}
We developed a stochastic-geometry framework for GA-WPIoT with spatially separated UAVs and annular GPB deployment. CSP moments and beta approximations characterize the
reliability distributions of energy transfer, uplink transmission, and their joint success. Numerical results show that the GPB inner radius and UAV altitude jointly affect reliability, with the preferred deployment depending on the propagation environment. In particular, a moderate GPB inner radius balances energy provisioning between central and edge devices, while UAV altitude balances the LoS probability gain against increased propagation loss. These findings provide guidelines for joint UAV--GPB
deployment to improve device-level reliability.
\vspace{-2.5mm}
\appendix
\section{Distance Distribution for the Annular GPB Deployment}
\label{app:nearest_gpb}
\vspace{-1.55mm}
This appendix provides the conditional distance distribution $F_{d_1^i}(r|v)$ and $f_{d_1^i}(r|v)$ used in the GPB-device-link analysis. Conditioned on $r_u^0=v$, the typical IoT device is placed at $\mathbf{z}_0=(v,0)$ without loss of generality. Let $D=d_1^i$ denote the distance from the typical IoT device to an arbitrary GPB. Since the GPBs are uniformly distributed in the annular region $\mathcal{A}_{\rm P}$, the conditional CDF of $D$ is
\begin{equation}
F_{d_1^i}(r|v)
=\mathbb{P}(D\leq r|r_u^0=v)=\frac{S(r,v)}{\pi(\hat R^2-\check R^2)},
\end{equation}
where$S(r,v)=|\mathcal{B}(\mathbf z_0,r)\cap\mathcal A_{\rm P}|$
denotes the intersection area between the IoT-centered disk and the
annular GPB deployment region.

According to the relative position between $\mathcal{B}(\mathbf{z}_0,r)$ and the annular boundaries,
$S(r,v)$ is given by
\begin{equation}
S(r,v)=
\begin{cases}
0, & \mathcal{E}_1,\\
\pi r^2, & \mathcal{E}_2,\\
\pi r^2-L(r,v;\check R), & \mathcal{E}_3,\\
L(r,v;\hat R), & \mathcal{E}_4,\\
L(r,v;\hat R)-L(r,v;\check R), & \mathcal{E}_5,
\end{cases}
\end{equation}
where $\mathcal{E}_1$--$\mathcal{E}_5$ represent no intersection, full inclusion in the annulus, intersection with only the inner boundary, intersection with only the outer boundary, and intersection with both boundaries, respectively.
The function $L(r,v;R)$ denotes the intersection area between two disks with radii $r$ and $R$ and center distance $v$, given by
\begin{equation}
\label{eq:lens_area}
{\footnotesize
\begin{aligned}
\mathcal{L}(r,v;R)&=\,r^2\cos^{-1}\left(\frac{v^2+r^2-R^2}{2vr}\right)+\,R^2\cos^{-1}\left(\frac{v^2+R^2-r^2}{2vR}\right)\\&-\,
\frac{1}{2}\sqrt{(-v+r+R)(v+r-R)(v-r+R)(v+r+R)}.\end{aligned}}
\end{equation}

Finally, the conditional PDF is obtained by
\begin{equation}
f_{d_1^i}(r|v)=\frac{\partial F_{d_1^i}(r|v)}{\partial r}.
\end{equation}

Accordingly, the nearest-GPB distance distribution used in
Section~\eqref{sec-PB_IoT_energy} is given by
\begin{align}
F_{d_1^0}(r|v)&=1-\left[1-F_{d_1^i}(r|v)\right]^{N_p}, \notag\\f_{d_1^0}(r|v)&=N_p\left[1-F_{d_1^i}(r|v)\right]^{N_p-1}f_{d_1^i}(r|v).
\end{align}

\bibliographystyle{IEEEtran}
\bibliography{references}

\end{document}